\documentclass[11pt]{article}

\usepackage[active]{srcltx}
\usepackage{hyperref}
\usepackage{amsmath,amssymb,amsthm}
\usepackage[all]{xy}

\renewcommand{\d}{\mathrm{d}}

\newtheorem{theorem}{Theorem}
\newtheorem{lemma}{Lemma}
\newtheorem{proposition}{Proposition}
\newtheorem{definition}{Definition}
\newcommand{\bd}{\begin{definicio} } 
\newcommand{\ed}{\end{definicio} } 
\newcommand{\bt}{\begin{theo} } 
\newcommand{\et}{\end{theo} } 
\newcommand{\bi}{\begin{itemize} } 
\newcommand{\ei}{\end{itemize} } 
\newcommand{\be}{\begin{enumerate} } 
\newcommand{\ee}{\end{enumerate} } 
\newcommand{\beq}{\begin{equation}}
\newcommand{\eeq}{\end{equation} } 
\newcommand{\br}{\begin{resultat} } 
\newcommand{\er}{\end{resultat} } 
\newcommand{\ba}{\begin{assum} } 
\newcommand{\ea}{\end{assum} } 
\newcommand{\bea}{\begin{eqnarray}}
\newcommand{\eea}{\end{eqnarray}}
\newcommand{\bal}{\begin{align*}}
\newcommand{\eal}{\end{align*}}
\newcommand{\bl}{\begin{lema}}
\newcommand{\el}{\end{lema}}
\newcommand{\bp}{\begin{prop}}
\newcommand{\ep}{\end{prop}}

\def\vf{\mathfrak X}

\def\Lag{{\cal L}}

\def\d{{\rm d}}

\def\Tan{{\rm T}}
\def\Lie{\mathop{\rm L}\nolimits}
\def\inn{\mathop{i}\nolimits}
\def\Cinfty{{\rm C}^\infty}

\def\beq{\begin{equation}}
\def\eeq{\end{equation}}
\def\bea{\begin{eqnarray}}
\def\eea{\end{eqnarray}}
\def\beann{\begin{eqnarray*}}
\def\eeann{\end{eqnarray*}}
\def\ben{\begin{enumerate}}
\def\een{\end{enumerate}}
\def\bit{\begin{itemize}}
\def\eit{\end{itemize}}

\def\tabaddress#1{{\small\it\begin{tabular}[t]{c}#1
\\[1.2ex]\end{tabular}}}

\title{\sc Order reduction, projectability and constraints of second-order field theories
and higher-order mechanics}

\author{
{\sc  Jordi Gaset},
   {\sc Narciso Rom\'an-Roy}  \\
   \tabaddress{Department of Mathematics.
   Ed. C-3, Campus Norte UPC\\
   C/ Jordi Girona 1. 08034 Barcelona. Spain.\\
   {{\bf e}-{\it mails}:
 \sl  gaset.jordi@gmail.com\ , narciso.roman@upc.edu}}}
   \date{\today \\
   }

\begin{document}

\maketitle

\pagestyle{myheadings}

\thispagestyle{empty}

\begin{abstract}
The projectability of Poincar\'e-Cartan forms
in a third-order jet bundle $J^3\pi$ onto a lower-order jet bundle is a consequence of the degenerate character of the corresponding Lagrangian.
This fact is analyzed using the constraint algorithm for the
associated Euler-Lagrange equations in $J^3\pi$.
The results are applied to study the Hilbert Lagrangian
for the Einstein equations (in vacuum)
from a multisymplectic point of view.
Thus we show how these equations are a consequence of the application of the constraint algorithm to the geometric field equations,
meanwhile the other constraints are related with the fact that
this second-order theory is equivalent to a first-order theory.
Furthermore, the case of higher-order mechanics 
is also studied as a particular situation.
\end{abstract}


\vbox{\raggedleft AMS s.\,c.\,(2010): \null  70H50,  53D42, 55R10, 83C05.}\null

\section{Introduction}
\label{intro}
There are some models in classical field theories
where, as a consequence of the singularity of the Lagrangian,
the order of the Euler-Lagrange equations is lower than expected.
A geometrical way of understanding this problem is
considering the projectability of the higher-order Poincar\'e-Cartan form
onto lower-order jet bundles \cite{first,Krupka,KrupkaStepanova,rosado,rosado2}.
We review the conditions for this projectability and study their consequences 
using the constraint algorithm
for the field equations of second order (singular) field theories,
thus enlarging the results stated in previous papers \cite{first,GcMM,KrupkaStepanova,rosado,rosado2}.
This constitutes the main result of the paper and it is stated in
Theorem \ref{prop:camps3}.

In this paper we restrict our study to second order field theories
in order to avoid some kinds of problems involving
the ambiguity in the definition of the
Poincar\'e-Cartan form in a higher-order jet bundle,
the non-uniqueness of the construction of the Legendre map
associated with a higher-order Lagrangian and the choice of the
multimomentum phase space for the Hamiltonian formalism
\cite{art:Aldaya_Azcarraga80,art:Francaviglia_Krupka82,
art:Kolar84,proc:Krupka84,Krupka}.
As it is well-known, for the second-order
case, all the Poincar\'e-Cartan forms are proved to be equivalent
and the Legendre map and the Hamiltonian multimomentum phase space
can be unambiguously defined \cite{pere,book:Saunders89,art:Saunders_Crampin90}.

As a relevant example, the case of the Hilbert Lagrangian for 
the Einstein equations with no matter fonts is analyzed.
In particular, we show how these equations are obtained as constraints appearing
as a consequence of the application of the constraint algorithm
to the geometric field equations which are stated in the
corresponding third-order jet bundle.
The other constraints arising in the algorithm are of 
geometrical nature. They are related with the fact that we are working with
some unnecessary degrees of freedom, because we are 
using a third-order jet bundle to describe a
second-order theory that, as a consequence of the 
projectibility of the Poincare-Cartan form, is really equivalent to a first-order theory  \cite{rosado2}.
In addition, this study constitutes a new approach to 
a multisymplectic formulation of the Lagrangian formalism for this model,
which is different to other previous attemps on this subject \cite{vey}.

Finally, this analysis is done
for the case of higher-order mechanics which,
as it is well-known, can be considered as a particular case of
higher-order field theories. Here we consider dynamical systems of any order,
since the above-mentioned ambiguities about the construction of the Poincar\'e-Cartan form
and the Legendre map do not occur in higher-order tangent bundles.

All the manifolds are real, second countable and $\Cinfty$. The maps and the structures are $\Cinfty$.  Sum over repeated indices is understood.
In order to use coordinate expressions,
remember that a multi-index $I$ is an element of $\mathbb{Z}^m$ where every
component is positive, the $i$th position of the multi-index is denoted $I(i)$, and
$|I| =\displaystyle\sum_{i=1}^{m} I(i)$ is the length of the multi-index.
An expression as $|I| = k$ means that the expression is taken for
every multi-index of length $k$.
Furthermore, the element $1_i\in\mathbb{Z}^m$ is defined as $1_i(j)=\delta_i^j$.
Finally, $n(ij)$ is a combinatorial factor which $n(ij)=1$ for $i=j$, and $n(ij)=2$ for $i\neq j$.

\section{Order reduction and projectability of the Poincar\'e-Cartan form}
\label{two}

Let $M$ be an \emph{m}-dimensional manifold and 
$\pi\colon E \rightarrow M$ a fiber bundle over M with $\dim E=m+n$
(the {\sl configuration bundle} of a classical field theory). 
The $k$-jet manifold of $\pi$ is denoted $J^k\pi$ and  is endowed with the natural projections 
$\pi^k_s\colon J^k\pi\rightarrow J^s\pi$,  
$\pi^k\colon J^k\pi\rightarrow E$, $\overline{\pi}^k\colon J^k\pi\rightarrow M$;
for $k>s\geq0$.
Then, a section $\psi\colon M\to J^k\pi$ of  $\bar{\pi}^{k}$
is {\sl holonomic} if $j^k(\pi^{k} \circ \psi) = \psi$; 
that is, $\psi$ is the $k$th prolongation of a section
$\phi = \pi^{k} \circ \psi\colon M\to E$.

Remember that a form $\omega\in\Omega^s(E)$ is said to be
 {\sl $\pi$-semibasic}  if \ $\inn(X)\omega=0$, and
{\sl $\pi$-basic} or {\sl $\pi$-projectable} if $\inn(X)\omega=0$ and  $\Lie(X)\omega=0$, 
for every $\pi$-vertical vector field $X\in \mathfrak{X}^V(\pi)$
(here, the symbols $\inn$ and $\Lie$ denote the inner contraction and the Lie derivative, respectively).
As a consequence of Cartan's formula,
$\Lie(X)\omega=\inn(X){\rm d}\omega + {\rm d}\inn(X)\omega$,
a form $\omega\in\Omega^n(E)$ is $\pi$-basic if, and only if, 
$\omega$ and ${\rm d}\omega$ are $\pi$-semibasic.

A special kind of vector fields are the {\sl coordinate total derivatives} 
\cite{pere,book:Saunders89}:
\beq
D_{i}=\frac{\partial}{\partial x^i}+\sum_{|I|=0}^k u_{I+1_i}^\alpha\frac{\partial}{\partial u_{I}^\alpha}\ .
\label{Di}
\eeq
For every function $f\in\Cinfty(J^k\pi)$, we have that
 $D_i f:=\Lie(D_i) f\in\Cinfty(J^{k+1}\pi)$. 
In addition, we have:
\bi{}
\item{}If $X\in\mathfrak{X}^V(\pi^k_s)$, then $[D_i,X]\in\mathfrak{X}^V(\pi^k_{s-1})$.
\item{}For $f\in C^{\infty}(J^{k}\pi)$, if $f$ is $\pi^k_s$-basic then $D_i f$ is $\pi^k_{s+1}$-basic.
\ei{}

We show some consequences of the projectability of the 
Poincar\'e-Cartan form for second order Lagrangian classical field theories. 
The {\sl Lagrangian form} that describes the theory is a $\overline{\pi}^2$-semibasic m-form 
$\mathcal{L}=L\,(\overline{\pi}^2)^*\omega\in \Omega^m(J^2\pi)$, 
where $L\in\Cinfty(J^2\pi)$ is the Lagrangian function,
$\omega$ is the volume form in $M$, 
and $\overline{\pi}^2\colon J^2\pi\to M$.
Natural coordinates of $J^3\pi$ adapted to the fibration are 
$(x^i,u^\alpha, u_i^\alpha,u_{I}^\alpha,u_{J}^\alpha)$, 
such that $\omega={\rm d} x^1\wedge\ldots\wedge{\rm d} x^m\equiv {\rm d}^mx$;
$1\leq i\leq m$, $1\leq \alpha \leq n$, and $I$, $J$ are multiindices 
with $|I|=2$, $|J|=3$, \cite{book:Saunders89}.

The Poincar\'e-Cartan $m$-form $\Theta_\mathcal{L}\in\Omega^m(J^{3}\pi)$ is locally given by
$$
\Theta_\mathcal{L}=L_\alpha^i{\rm d}u^\alpha \wedge{\rm d}^{m-1}x_i+L_\alpha^{ij}{\rm d}u_i^\alpha \wedge{\rm d}^{m-1}x_j +\left(L- L_\alpha^i u_i^\alpha - L_\alpha^{ij}u_{1_i+1_j}^\alpha \right){\rm d}^mx \ ,
$$
where $\displaystyle {\rm d}^{m-1}{x_j}=\inn\left({\frac{\partial}{\partial x^j}}\right){\rm d}^mx$
and the functions $L_\alpha^i,L_\alpha^{ij}\in C^{\infty}(J^{3}\pi)$ are
$$
L_\alpha^i=\frac{\partial L}{\partial u_{i}^\alpha} - D_j L^{ij}_{\alpha}\quad ;\quad L_\alpha^{ij}=\frac{1}{n(ij)}\frac{\partial L}{\partial u_{1_i+1_j}^\alpha} \ .
$$

\begin{lemma}
\label{lema:camps}
For $s=1,2$, the following conditions are equivalent:
\begin{enumerate}
\item $\Theta_\mathcal{L}$ projects onto $J^s\pi$. 
\item
$\rm d\Theta_\mathcal{L}$ is $\pi^3_s$-semibasic.
\item
$\Lie(X) L^i_\alpha=0$ and $\Lie(X) L^{ij}_\alpha=0$; 
for every $X\in\mathfrak{X}^V(\pi^3_s)$.
\end{enumerate}
\end{lemma}
({\sl Proof\/}):\quad
 $({\rm 1}\Leftrightarrow{\rm 2}$) is a consequence of Cartan's formula. 

For (${\rm 2}\Leftrightarrow{\rm 3}$), in the case $s=2$, 
we compute the condition {\rm 2} in coordinates. 
It turns to be equivalent to
$$
\frac{\partial L_\alpha^i}{\partial u^\beta_J}=0
\quad , \quad
\frac{\partial L_\alpha^{ij}}{\partial u^\beta_J}=0
\quad , \quad
\frac{\partial}{\partial u^\beta_J}(L- L_\alpha^i u_i^\alpha - L_\alpha^{ij}u_{1_i+1_j}^\alpha)=0 \quad ;
$$ 
(for $|J|=3$, and for every $\beta$, $\alpha$, $i$ and $j$). 
The last equation is a consequence of the other two
(because $L$ does not depend on $u^\beta_J$);
which are locally equivalent to {\rm 3}, since
$\displaystyle \left\{ \frac{\partial}{\partial u^\beta_J}\right\}$
generates $\mathfrak{X}^V(\pi^3_s)$. 
The case $s=1$ can be proved in a similar way.
\qed 

Other important results concerning to this topic
(that we present here for completeness) 
are the following \cite{rosado}:

\begin{proposition}
\label{prop:camps1}
If  $\Theta_\mathcal{L}$ projects onto $J^s\pi$, then the order of the Euler-Lagrange equations is at most $s+1$.
\end{proposition}

\begin{proposition}\label{prop:camps2}
If there exist $\mathcal{L'}\in \Omega^m(J^{1}\pi)$ such that $\Theta_\mathcal{L}=(\pi^3_{1})^* \Theta_\mathcal{L'}$, then $\mathcal{L}=(\pi^3_{1})^*\mathcal{L'}$.
\end{proposition}

Concerning to the last proposition, the study of the existence of 
an equivalent lower order Lagrangian $\mathcal{L'}\in \Omega^m(J^{1}\pi)$
has been analysed in \cite{first, rosado2}.

If the Poincar\'e-Cartan form $\Theta_\mathcal{L}$ projects onto 
a lower-order jet bundle, it is associated to a highly degenerate Lagrangian
(this is just a consequence of the third item in Lemma \ref{lema:camps}).
As a consequence of this fact, the field equations could not have admissible solutions
everywhere in $J^3\pi$, but in some submanifold of it which can be obtained after applying
a suitable constraint algorithm (see, for instance, \cite{art:deLeon_Marin_Marrero_Munoz_Roman05}).

In order to study these facts, we introduce the following concepts
\cite{art:Echeverria_Munoz_Roman98}:

\begin{definition}
An $m$-{\rm multivector field}  in $J^3\pi$ is a skew-symmetric contravariant 
tensor of order $m$ in $J^3\pi$. The set of $m$-multivector fields 
in $J^3\pi$ is denoted $\vf^m (J^3\pi)$.

A multivector field $\mathbf{X}\in\vf^m(J^3\pi)$ is said to be {\rm locally decomposable} if,
for every $p\in J^3\pi$, there is an open neighbourhood  $U_p\subset J^3\pi$
and $X_1,\ldots ,X_m\in\vf (U_p)$ such that $\mathbf{X}\vert_{U_p}=X_1\wedge\ldots\wedge X_m$.

Non-vanishing locally decomposable $m$-multivector fields $\mathbf{X}\in\vf^m(J^3\pi)$ are locally associated with $m$-dimensional
distributions $D\subset\Tan J^3\pi$. Then,
$\mathbf{X}$ is {\rm integrable} if its associated distribution is integrable. 
In particular,
$\mathbf{X}$ is {\rm holonomic} if
it is integrable and 
its integral sections are holonomic sections of $\bar\pi^3$.
\end{definition}

Then, the solutions to the Euler-Lagrange equations for a
second-order field theory are
the integral sections of locally decomposable holonomic multivector fields
${\bf X}\in\mathfrak{X}^m(J^3\pi)$ such that 
\beq 
\label{eq:fields1}
\inn({\bf X}){\rm d}\Theta_\Lag=0 \ .
\eeq
Therefore:

\begin{theorem}
\label{prop:camps3}
If $\Theta_\mathcal{L}$ projects onto $J^s\pi$, then solutions to 
the corresponding Euler-Lagrange equations only exist in the points of a submanifold
$\mathcal{S}\hookrightarrow J^3\pi$, where $\mathcal{S}$ is
locally defined  by the constraint functions given by
\begin{itemize}
\item
$L_\alpha^0=0$; if $s=2$.
\item
$L_\alpha^0=0$ and
$\displaystyle D_iL_\alpha^0=0$; if $s=1$.
\end{itemize}
Where \ $\displaystyle L_\alpha^0=\frac{\partial L}{\partial u^\alpha}-D_iL^i_\alpha=\frac{\partial L}{\partial u^\alpha}-D_i\frac{\partial L}{\partial u^\alpha_i}+D_I\frac{\partial L}{\partial u^\alpha_I}$\ .
\end{theorem}
({\sl Proof\/}):\quad 
${\bf X}$ can be written in coordinates as 
$$
{\bf X}=f\bigwedge_{i=1}^m\left(D_i+(F_{J,i}^\alpha-u_{J+1_i}^\alpha) \frac{\partial}{\partial u_J^\alpha}\right)=f\bigwedge_{i=1}^m X_i \ ;
$$
for $f,F^\alpha_{J,i}\in C^\infty(J^3\pi)$, ($|J|=3$).
Using this expression, equation (\ref{eq:fields1}) reduces to 
\beq
L^0_\alpha+(F_{J,i}^\beta-u_{J+1_i}^\beta)\frac{\partial L^i_\alpha}{\partial u^\beta_J}=0 \ ,
\label{ELmvf}
\eeq
 which are the Euler-Lagrange equations for multivector fields. If $\Theta_\mathcal{L}$ projects either onto $J^1\pi$ or $J^2\pi$,
by Lemma \ref{lema:camps} we have $\displaystyle\frac{\partial L^i_\alpha}{\partial u^\beta_J}=0$, and then
from \eqref{ELmvf} we get $L_\alpha^0=0$.
Observe that, as a consequence, we cannot compute any of the functions 
$F^\alpha_{J,i}$. 
Actually $L_\alpha^0=0$ are restrictions for the points of the manifold $J^3\pi$,
which we assume that define a submanifold ${\cal S}_1\subset J^3\pi$,
where the equation \eqref{eq:fields1} have solutions.
In order to find $F^\alpha_{J,i}$ we use the constraint algorithm 
(as it is outlined, for instance, in \cite{pere}).
So we look for the points of ${\cal S}_1$ where the multivector fields which are solutions to \eqref{eq:fields1}
(on ${\cal S}_1$) are tangent to ${\cal S}_1$.
Thus, imposing this {\sl consistency} or {\sl tangency condition} we get
$$
0=\Lie({X_i})L_\alpha^0=D_iL^0_\alpha+\left(F^\beta_{J,i}-u_{J+1_i}^\beta\right)\frac{\partial L^0_\alpha}{\partial u^\beta_{J}} \qquad \mbox{\rm (on ${\cal S}_1$)}\ .
$$
If $\Theta_\mathcal{L}$ projects onto $J^1\pi$, 
then the associated Euler-Lagrange equations are of order at most 2 
(by proposition \ref{prop:camps1}). This implies that $L^0_\alpha$, 
which are the Euler-Lagrange equations before being evaluatedon sections, 
are $\pi^3_2$-projectable. Thus, 
$\displaystyle\frac{\partial L^0_\alpha}{\partial u^\beta_{J}}=0$, 
and we find new restrictions, $D_iL^0_\alpha=0$
which are assumed to define a new submanifold 
${\cal S}_2\subset{\cal S}_1\subset J^3\pi$ where the 
solutions to \eqref{eq:fields1} are tangent to ${\cal S}_1$.
\qed

Notice that, depending on the Lagrangian, we may need to continue 
the constraint algorithm, so obtaining that
$$
D_jD_iL^0_\alpha+\left(F^\beta_{J,j}-u_{J+1_j}^\beta\right)\frac{\partial D_iL^0_\alpha}{\partial u^\beta_{J}}=0  \qquad \mbox{\rm (on ${\cal S}_2$)} \quad .
$$
This process continues until the new conditions hold identically
and we find a {\sl final constraint submanifold} 
${\cal S}_f$ of $J^3\pi$ where solutions to
\eqref{eq:fields1} are tangent to ${\cal S}_f$.

\section{The Hilbert-Einstein Lagrangian}

Here $M$ is a $4$-manifold representing space-time and the fibers are the spaces of Lorentzian metrics. 
The fiber coordinates in $E$ are $(x^\mu,g_{\mu\nu})$ ($\mu,\nu$ and
all greek indices in this section run from $0$ to $3$),
where $g_{\mu\nu}$ are the component functions of the metric.
The Hilbert Lagrangian function without matter is:
$$
L=\sqrt{|{\rm det}(g)|}\,R=\sqrt{|{\rm det}(g)|}\,g^{\mu\nu}R_{\mu\nu} \ ,
$$
where $R=g^{\mu\nu}R_{\mu\nu}$ is the {\sl scalar curvature},
$R_{\mu\nu}=D_\rho\Gamma^{\rho}_{\mu\nu}-D_\mu\Gamma^{\rho}_{\rho\nu}+
\Gamma^{\rho}_{\mu\nu}\Gamma^{\delta}_{\delta\rho}-
\Gamma^{\rho}_{\delta\nu}\Gamma^{\delta}_{\mu\rho}$
are the components of the {\sl Ricci tensor},
$\displaystyle\Gamma^{\rho}_{\mu\nu}=
\frac{1}{2}\,g^{\rho\lambda}\left(\frac{\partial g_{\nu\lambda}}{\partial x^\mu}+ 
\frac{\partial g_{\lambda\mu}}{\partial x^\nu}- \frac{\partial g_{\mu\nu}}{\partial x^\lambda}\right)$ 
are the {\sl Christoffel symbols of the Levi-Civita connection} of $g$, and
$g^{\mu\nu}$ denotes the inverse matrix of $g$, 
namely: $g^{\mu\nu}g_{\nu\rho}=\delta^\mu_\rho$.
As the Christoffel symbols depend on first-order derivatives of $g_{\mu\nu}$ and
taking into account the expression \eqref{Di} we have that
$R$ contains second-order derivatives of the components of the metric
and thus this is a second-order field theory. 

The Poincar\'e-Cartan form $\Theta_\mathcal{L}$ 
associated with the Hilbert Lagrangian density 
$\mathcal{L}=L\,(\overline{\pi}^2)^*\omega=L\,\d^4x$ is
\beann
&\Theta_\mathcal{L}&=-\left(\sum_{\alpha\leq\beta}L^{\alpha\beta,\mu}g_{\alpha\beta,\mu}+\sum_{\alpha\leq\beta}L^{\alpha\beta,I}g_{\alpha\beta,I}-\sum_{\alpha\leq\beta}L\right)\d^4x \nonumber \\
&&+\sum_{\alpha\leq\beta}L^{\alpha\beta,\mu}\d g_{\alpha\beta}\wedge \d^{m-1}x_\mu+\sum_{\alpha\leq\beta}L^{\alpha\beta,\mu\nu}\d g_{\alpha\beta,\mu}\wedge \d^{m-1}x_{\nu}  \ ;
\eeann
where
\bea
\label{1eq}
L^{\alpha\beta,\mu}&=&\frac{\partial L}{\partial g_{\alpha\beta,\mu}} - \sum_{\nu=0}^{3}\frac{1}{n(\mu\nu)}D_\nu\left( \frac{\partial L}{\partial g_{\alpha\beta,\mu\nu}}\right)
\\
\nonumber
&=&\frac{ n(\alpha\beta)}{2}\,\sqrt{|{\rm det}(g)|}\left( \Gamma^\alpha_{\nu\sigma}(g^{\beta\sigma}g^{\mu\nu}-g^{\beta\mu}g^{\sigma\nu})+\Gamma^\beta_{\nu\sigma}(g^{\alpha\sigma}g^{\mu\nu}-g^{\alpha\mu}g^{\sigma\nu})\right)
\\
\label{2eq}
L^{\alpha\beta,\mu\nu}&=&\frac{1}{n(\mu\nu)}\frac{\partial L}{\partial g_{\alpha\beta,\mu\nu}}=\frac{n(\alpha\beta)}{2}\,\sqrt{|{\rm det}(g)|}(g^{\alpha\mu}g^{\beta\nu}+g^{\alpha\nu}g^{\beta\mu}-2g^{\alpha\beta}g^{\mu\nu}) \ .
\eea
This form projects onto $J^1\pi$
and hence the propositions of section 2 hold. As is well known, the
corresponding Euler-Lagrange equations, which are
essentially the Einstein equations \cite{Carroll}, are of second order. 

Moreover, as it is noted in \cite{rosado}, 
the projected  form $(\pi^3_1)^*\Theta_{\mathcal{L}}$ 
is not the Poincar\'e-Cartan form of any Lagrangian of order 1. 
Nevertheless, there exists a Lagrangian of order 1 
whose Euler-Lagrange equations have solutions which are the same than
those for the Hilbert Lagrangian \cite{first,rosado2}.

Finally, we apply in detail the theorem \ref{prop:camps3} to the Hilbert Lagrangian (that is, the constraint algorithm). The local expression of a holonomic and locally decomposable multivector field in $J^3\pi$ is
$$
{\bf X}_\Lag=\bigwedge_{\rho=0}^3X_\rho=
\bigwedge_{\rho=0}^3\left(D_\rho+\sum_{\alpha\leq\beta}(F_{\alpha\beta;J,\rho}-g_{\alpha\beta;J+1_\rho})\frac{\partial}{\partial g_{\alpha\beta;J}}\right) \ ,
$$
and the equations \eqref{eq:fields1} take the local expression:
\bea
D_\mu L^{\alpha\beta,\mu}-\frac{\partial L}{\partial g_{\alpha\beta}}&=&0
\label{1} \\
D_\nu L^{\alpha\beta,\mu\nu} +L^{\alpha\beta,\mu}-\frac{\partial L}{\partial g_{\alpha\beta,\mu}}&=&0
\label{2}  \\
n(\mu\nu)n(\alpha\beta)L^{\alpha\beta,\mu\nu}-n(\alpha\beta)\frac{\partial L}{\partial g_{\alpha\beta,\mu\nu}}&=&0 \ .
\label{3} 
\eea
The equations \eqref{2} and \eqref{3} are just the
identities \eqref{1eq} and \eqref{2eq}. 
Furthermore, using \eqref{1eq} we see that equations \eqref{1} are:
\beq
0  =\frac{\partial L}{\partial g_{\alpha\beta}}-D_\mu\frac{\partial L}{\partial g_{\alpha\beta,\mu}}+D_I\frac{\partial L}{\partial g_{\alpha\beta,I}}=
-\,\sqrt{|{\rm det}(g)|}\,n(\alpha\beta) \left(R^{\alpha\beta}-\frac{1}{2}g^{\alpha\beta}R\right)
\equiv L^{\alpha\beta} \ .
\label{Eeq}
\eeq
Notice that with these equations we cannot determine any of the unknows
$F_{\alpha\beta;J,\rho}$. Actually
$L^{\alpha\beta}$ project onto ${J^2\pi}$; hence they do not depend on the higher-order derivatives and
therefore $L^{\alpha\beta}=0$ are constraints
which define the submanifold $\mathcal{S}_1\subset J^3\pi$.
These functions, evaluated on the points of holonomic sections of $\bar\pi^3$
are the Euler-Lagrange equations; that is,
they give the Einstein equations 
$$
R^{\alpha\beta}-\frac{1}{2}g^{\alpha\beta}R=0 \ ,
$$
which, 
in this way, turn out to be constraints defining the submanifold $\mathcal{S}_1$.
The tangency conditions for these functions
$L^{\alpha\beta}$ lead to
\beq \label{eq4}
\Lie({X_\rho})L^{\alpha\beta}=D_\rho L^{\alpha\beta}+
\sum_{\mu\leq\nu}(F_{\mu\nu;J,\rho}-
g_{\mu\nu;J+1_\rho})\frac{\partial L^{\alpha\beta}}{\partial g_{\mu\nu;J}} =D_\rho L^{\alpha\beta}=0
\quad \mbox{ (on ${\cal S}_1$)} \ ,
\eeq
since $\displaystyle \frac{\partial L^{\alpha\beta}}{\partial g_{\mu\nu;J}}=0$.
By the properties of the total derivative, we have that the functions $D_\rho L^{\alpha\beta}$ project onto $J^3\pi$ and then
the functions $D_\rho L^{\alpha\beta}$ are constraints again 
and define the submanifold ${\cal S}_2\subset{\cal S}_1\subset J^3\pi$
(this is also obvious bearing in mind \eqref{Di}). 
Finally, the new tangency conditions lead to the equalities
\beq
D_\tau D_\rho L^{\alpha\beta}+\sum_{\mu\leq\nu}(F_{\mu\nu;J,\tau}-g_{\mu\nu;J+1_\tau})\frac{\partial D_\rho L^{\alpha\beta}}{\partial g_{\mu\nu;J}}=0
\quad \mbox{ (on ${\cal S}_2$)} \ ,
\label{Eq:LastConsAlgo}
\eeq
which are not constraints since they contain the unknown functions $F_{\mu\nu;J,\tau}$.

In order to understand the implications of 
equations \eqref{eq4} and \eqref{Eq:LastConsAlgo}, 
consider an holonomic section 
$\psi\colon M\rightarrow J^3\pi$. 
When evaluated at the section, they look:
$$
\nonumber\left(D_\rho L^{\alpha\beta}\right)\Big\vert_{\psi}=\frac{\partial\left(L^{\alpha\beta}\circ{\psi}\right)}{\partial x^\rho}=0
\quad , \quad
\nonumber\left(D_\tau D_\rho L^{\alpha\beta}\right)\Big\vert_{\psi}=\frac{\partial^2\left(L^{\alpha\beta}\circ{\psi}\right)}{\partial x^\tau\partial x^\rho}=0 \ .
$$
Here we have used that,
if $\psi$ is an integral section of $\bf{X}$, 
then $(F_{\mu\nu;J,\tau}-g_{\mu\nu;J+1_\tau})|_\psi=0$.
So, if $\psi$ is a solution to the Einstein equations (that is, $L^{\alpha\beta}\circ{\psi}=0$), then $\psi$ also satisfies equations \eqref{eq4} and \eqref{Eq:LastConsAlgo}.
Therefore, from the physical point of view,
the only relevant equations are \eqref{Eeq},
which are equivalent to the Einstein equations.
The other equations \eqref{eq4} and \eqref{Eq:LastConsAlgo}
contain no physical information: they are of 
geometrical nature. They arise from the fact that 
we are using a third-order jet bundle $J^3\pi$, prepared for
describing a second-order theory,
for a Lagrangian which is physically equivalent to 
a first-order Lagrangian and hence,
we have redundant information.

In a further paper, Hilbert's Lagrangian as well as other Lagrangian models
for gravitation will be studied in detail using this procedure and the unified formalism developed in \cite{pere}.

\section{Application to higher-order mechanics}

Now, consider the particular case where $\pi\colon E \rightarrow \mathbb{R}$,
with $\dim E=n+1$, is the {\sl configuration bundle} of a higher-order non-autonomous theory.
We have the natural projections 
$\pi^k_s\colon J^k\pi\rightarrow J^s\pi$,  
$\pi^k\colon J^k\pi\rightarrow E$, $\overline{\pi}^k\colon J^k\pi\rightarrow M$;
for $k>s\geq0$.
As above, natural coordinates in $J^{2k-1}\pi$ are
$(t,q_i^\alpha)$; $0\leq i\leq 2k-1$, $1\leq\alpha\leq n$.
The (only) {\sl total time derivative} is
$$
D_t=\frac{\partial}{\partial t}+\sum_{i=0}^k q_{i+1}^\alpha\frac{\partial}{\partial q_i^\alpha}\ ,
$$
which verifies the properties stated in Section \ref{two}.
The dynamics is given by a {\sl Lagrangian form}  $\mathcal{L}\in \Omega^1(J^k\pi)$, which is
a $\overline{\pi}^k$-semibasic 1-form and
it has associated the Lagrangian function  $L\in\Cinfty(J^k\pi)$, such that
$\mathcal{L}=L\,(\overline{\pi}^k)^*\d t$, where
$\d t$ is the canonical volume form in $\mathbb{R}$ \cite{deLeon}. 
The Poincar\'e-Cartan 1-form $\Theta_\mathcal{L}\in\Omega^1(J^{2k-1}\pi)$ is given  locally by:
$$
\Theta_\mathcal{L}=\sum_{r=1}^k L_\alpha^r \d q_{r-1}^\alpha+\left(L-\sum_{r=1}^k L_\alpha^r q_r^\alpha \right)\d t \ ,
$$
where the functions $L_\alpha^r\in C^{\infty}(J^{2k-1}\pi)$ are
$$
L_\alpha^r=\sum_{i=0}^{k-r}(-1)^iD_t^i\left( \frac{\partial L}{\partial q_{r+i}^\alpha}\right) \ ,
$$
and they can be obtained inductively by setting $L_\alpha^r=0$, for $r>k$, and
$$
L_\alpha^r=\frac{\partial L}{\partial q_{r}^\alpha}-D_tL^{r+1}_\alpha \ .
$$

The properties stated in Lemma \ref{lema:camps} and
Propositions \ref{prop:camps1} and \ref{prop:camps2} read:

\begin{lemma}
\label{lema:ho}
For $s\geq k-1$, the following conditions are equivalent:
\begin{enumerate}
    \item
$\Theta_\mathcal{L}$ projects onto $J^s\pi$.
    \item
$\d\Theta_\mathcal{L}$ is $\pi^{2k-1}_s$-semibasic.
    \item
$\Lie(X) L^r_\alpha=0$; for every $X\in\mathfrak{X}^V(\pi^{2k-1}_s)$,
and for $r=1,\dots,k$, $\alpha=1,\dots,n$.
\end{enumerate}
\end{lemma}
({\sl Proof\/}):\quad
$({\rm 1}\Leftrightarrow{\rm 2}$) is a consequence of Cartan's formula. For the equivalence between $2$ and $3$ we consider two cases:

\noindent - If $s\geq k$: The relevant terms of 
$\d\Theta_\mathcal{L}$ are of the form:
$$
\frac{\partial L_\alpha^i}{\partial q_r^\beta}\d q_r^\beta\wedge\d q_{i-1}^\alpha \quad , \quad 
\frac{\partial}{\partial q_r^\beta}\left(L-\sum_{i=1}^k L_\alpha^i q_i^\alpha \right)\d q_r^\beta\wedge\d t\quad ; \quad
s<r\leq 2k-1 \ .
$$
Then, $\d\Theta_\mathcal{L}$ is $\pi^{2k-1}_s$-semibasic if, and only if,
$\displaystyle \frac{\partial L_\alpha^i}{\partial q_r^\beta}=0$,
and this is equivalent to $\Lie(X) L^r_\alpha=0$, 
for every $X\in\mathfrak{X}^V(\pi^{2k-1}_s)$,
since $\displaystyle \left\{\frac{\partial}{\partial q_r^\beta}\right\}$
generates $\vf^V(\pi^{2k-1}_s)$.

\noindent - If $s=k-1$: In this case 
$\d\Theta_\mathcal{L}$ is $\pi^{2k-1}_s$-semibasic if, and only if, 
$$\frac{\partial L_\alpha^i}{\partial q_r^\beta}=0 \quad , \quad
\frac{\partial L}{\partial q_k^\beta}-L_\beta^k=0 \quad ;
$$
 but this last condition is fulfilled by the definition of $L_\beta^k$, and the same reasoning above allows us to prove the statement.
\qed

If $\Theta_\mathcal{L}$ projects onto $J^s\pi$, with $s<k-1$, 
then $L$ does not depend on $q^\alpha_j$, for $j>s+1$,
then there exists a function $L'\in C^\infty(J^{s+1}\pi)$ such that 
$L=(\pi^{k}_{s+1})^*L'$ and the theory is not strictly of order $k$.
Furthermore, in the case $s\geq k-1$, a Lagrangian such that 
$\Theta_\mathcal{L}$ projects onto $J^s\pi$ depends on all the variables
and thus we have a theory of order $k$,
although the associated Euler-Lagrange equations 
are of lower order as a system of differential equations. In fact:

\begin{proposition}
\label{prop:ho1}
If  $\Theta_\mathcal{L}$ projects onto $J^s\pi$, then the order of the Euler-Lagrange equations is at most $s+1$.
\end{proposition}
({\sl Proof\/}):\quad
Note that $L_\alpha^0\in C^{\infty}(J^{2k}\pi)$.
 For a curve $\phi\in\Gamma(\pi)$ which is a solution to the Euler-Lagrange equations
we have that $L_\alpha^0|_{j^{2k-1}\phi}=0$.  Then, 
for $X\in\mathfrak{X}^V(\pi^k_{s+1})$,
$$
\Lie(X)L_\alpha^0=\Lie(X)\frac{\partial L}{\partial q_{0}^\alpha}-\Lie(X)(D_tL^{1}_\alpha)=
\Lie(X)\frac{\partial L}{\partial q_{0}^\alpha}-D_t(\Lie(X)L^{1}_\alpha)-\Lie({[X,D_t]})L^{1}_\alpha \ .
$$
Since $[D_t,X]\in\mathfrak{X}^V(\pi^k_{s})$ and $L^{1}_\alpha$ and $L$ are $\pi^k_s$-basic,
then $\Lie(X)(L_\alpha^0)=0$. Therefore, after evaluating on the section, the resulting equations only contain derivations up to order $s+1$.
\qed

Equating the local expressions of  $\Theta_\mathcal{L}$ and $\Theta_\mathcal{L'}$
the following result holds immediately:

\begin{proposition}
\label{prop:ho2}
If there exist $\mathcal{L}'\in\Omega^{1}(J^{k'}\pi)$ such that $\Theta_\mathcal{L}=(\pi^{2k-1}_s)^* \Theta_\mathcal{L'}$, then $\mathcal{L}=(\pi^{2k-1}_s)^*\mathcal{L'}$.
\end{proposition}

In particular $L$ is not strictly of order $k$.

Finally, a similar result to theorem \ref{prop:camps3} is the following:

\begin{theorem}
\label{prop:ho3}
If $\Theta_\mathcal{L}$ projects onto $J^s\pi$, then 
solutions to the corresponding Euler-Lagrange equations  exist
only in points of a submanifold
$\mathcal{S}\hookrightarrow J^{2k-1}\pi$, where $\mathcal{S}$ is locally defined by the constraint functions given by 
\begin{equation}\nonumber
D_t^jL_\alpha^0=0\quad ;\quad (j=0,\dots,2k-s-2) \ .
\end{equation}
\end{theorem}
({\sl Proof\/}):\quad
To find a solution to the Euler-Lagrange equations is equivalent to find a holonomic vector field $X\in\mathfrak{X}(J^{2k-1}\pi)$ such that 
\beq \label{eq:ho1}
\inn(X)\d\Theta_\Lag=0 \ .
\eeq
The holonomic vector fields have the local expression:
$$
X=D_t+\left(F^\alpha-q_{2k}^\alpha\right)\frac{\partial}{\partial q^\alpha_{2k-1}} \ ,
$$
and then equation (\ref{eq:ho1}) reduces to 
$$
L^0_\alpha-(F^\beta-q_{2k}^\beta)\frac{\partial L^1_\alpha}{\partial q^\beta_{2k-1}}=0 \ .
$$
If $\Theta_\mathcal{L}$ projects onto $J^s\pi$ for $s<2k-1$, the second term vanishes
and $L_\alpha^0=0$.
Notice that we cannot compute any function $F^\alpha$. 
Actually  $L_\alpha^0\in C^\infty(J^{2k-1}\pi)$, thus  $L_\alpha^0=0$ 
is just a restriction for the points of the manifold $J^{2k-1}\pi$.
Next, following the constraint algorithm \cite{constraint}, we
impose the tangency condition and we get
$$
0=\Lie(X)L_\alpha^0=D_tL^0_\alpha+\left(F^\alpha-q_{2k}^\alpha\right)\frac{\partial L^0_\alpha}{\partial q^\alpha_{2k-1}}\ .
$$
If $\Theta_\mathcal{L}$ projects onto $J^s\pi$,
 then the second term vanishes (Proposition \ref{prop:ho1}) 
and we find another constraint, $D_tL^0_\alpha=0$.
The algorithm continues until we reach the condition
$D_t^{2k-s-2}L_\alpha^0=0$.
\qed

As above, depending on the Lagrangian, we may need to continue 
the constraint algorithm, obtaining that
$$
0=D_t\left(D_t^{2k-s-2}L^0_\alpha\right)+\left(F^\alpha-q_{2k}^\alpha\right)\frac{\partial}{\partial q^\alpha_{2k-1}}\left(D_t^{2k-s-2}(L^0_\alpha)\right) \ .
$$
This process continues until the new conditions hold identically.


\section*{Acknowledgments}

We acknowledge the financial support of the 
{\sl Ministerio de Ciencia e Innovaci\'on} (Spain), project 
MTM2014--54855--P, and of
{\sl Generalitat de Catalunya}, project 2014-SGR-634.
We want to thank to the referees for their valuable comments and suggestions that
have allowed us to improve the final version of this work.

\end{document}